# DeepMCDose: A Deep Learning Method for Efficient Monte Carlo Beamlet Dose Calculation by Predictive Denoising in MR-Guided Radiotherapy

Ryan Neph[1][0000-0002-8176-6723], Yangsibo Huang[1], Youming Yang[1], and Ke Sheng[1]

[1] University of California – Los Angeles, Los Angeles, CA 90095, USA
ryanneph@ucla.edu

**Abstract.** The next great leap toward improving treatment of cancer with radiation will require the combined use of online adaptive and magnetic resonance guided radiation therapy techniques with automatic X-ray beam orientation selection. Unfortunately, by uniting these advancements, we are met with a substantial expansion in the required dose information and consequential increase to the overall computational time imposed during radiation treatment planning, which cannot be handled by existing techniques for accelerating Monte Carlo dose calculation. We propose a deep convolutional neural network approach that unlocks new levels of acceleration and accuracy with regards to post-processed Monte Carlo dose results by relying on data-driven learned representations of low-level beamlet dose distributions instead of more limited filter-based denoising techniques that only utilize the information in a single dose input. Our method uses parallel UNET branches acting on three input channels before mixing latent understanding to produce noise-free dose predictions. Our model achieves a normalized mean absolute error of only 0.106% compared with the ground truth dose contrasting the 25.7% error of the under sampled MC dose fed into the network at prediction time. Our model's per-beamlet prediction time is ~220ms, including Monte Carlo simulation and network prediction, with substantial additional acceleration expected from batched processing and combination with existing Monte Carlo acceleration techniques. Our method shows promise toward enabling clinical practice of advanced treatment technologies.

**Keywords:** Radiation Dose Prediction, Deep Learning, CNN, Monte Carlo.

## 1 Introduction

Magnetic resonance guided radiotherapy (MRgRT) is an innovation that asserts dominance over traditional CT-guided radiotherapy with respect to the offered soft tissue contrast and imaging flexibility. Such innovations in the pre-treatment imaging and the online image-guided contexts have enabled enhanced precision in the treatment of inconspicuous and moving lesions. The difficulty of widespread adoption of MRgRT is in part due to the complicating behavior of charged particles (electrons) in the presence of a moderate to strong magnetic field. The result is a non-negligible perturbation to the more typical dose distributions observed without a strong magnetic field. Great effort has been invested in acceleration of deterministic dose calculation, including the works of Chen [1], Neylon [2] and most recently Neph [3] which emphasize efficient GPU implementation. However, the effects of a strong magnetic field fundamentally invalidate the assumptions made by these heavily relied upon deterministic dose calculation algorithms, leaving us instead with highly the accurate and flexible, but comparably less efficient Monte Carlo (MC) dose simulation technique.

The intersection of MRgRT with other advanced clinical techniques presents a serious challenge with respect to the capabilities of existing MC dose calculation tools. Online adaptive radiotherapy (OART) deviates from the clinical standard by both re-imaging and re-optimizing RT treatment plans prior to each daily radiation delivery. The outcome of OART is increased delivery precision and improved patient outcome but is commonly rendered computationally intractable given the insufficient speed of both the dose calculation and plan optimization stages. Additionally, the innovation of beam orientation optimization (BOO) increases plan quality while simplifying the planning effort. However, BOO imparts a substantial requirement on the compulsory dose data that is calculated prior to the start of planning. Current clinical practice with human pre-selection of around 10 static beams necessitates calculation of *planning* dose distributions for a few thousand individual beamlets. By comparison, joint optimization of beam orientations and their fluence maps performed by 4pi treatment planning, considers 1162 candidate beam orientations and requires calculation of dose for hundreds of thousands of beamlets consequently.

It is well understood that each of these techniques offer significant and complementary improvements to the treatment planning process and quality of patient care. However, the convergence of their practice imposes formidable challenges on the dose calculation component of the planning process; namely that we must simultaneously pivot to using more accurate methods, which can handle EREs, while greatly increasing the efficiency to handle significant increases to the amount of prerequisite data. In summary, we must find a way to get the accuracy benefits of MC dose simulation while accelerating its computation time beyond that which is possible using any existing MC acceleration techniques.

Previous work on accelerating MC simulation has investigated the use of denoising algorithms applied to under sampled (noisy) MC dose. Deasy [4] used a wavelet coefficient thresholding approach to denoising on a slice-by-slice basis. Kawrakow [5] presents a 3D implementation of locally adaptive Savitzky-Golay filtering that selects the anisotropic filter window size by means of a locally supported chi-square test, limiting the effect of systematic bias. Fippel [6] proposed an optimization approach including both dose fidelity and smoothness regularization terms. Miao [7] investigated the use of an adaptive denoising approach modeling the dose in terms of heat transport and used anisotropic diffusion to achieve smoothed distributions. El Naqa [8] used a hybrid median filtering approach which adapts the filter to the local content of the dose distribution to more effectively tradeoff the benefits of mean- and median-based denoising.

This existing work places an emphasis on only moderately under sampled dose suggesting their incapacity to robustly and accurately denoise dose with anything beyond this modest level of noise or in heterogeneous geometries. El Naqa [9] judges that "uncertainties of greater than 5% are probably too large" for producing clinically usable treatment plans, and that "maximum error of denoised distributions can still be large for raw MC uncertainties of 3%", indicating observed errors up to 15% in these cases.

Our contributions focus on meeting this need. We harness a successful Deep Learning model architecture, UNET [10], to perform concurrent denoising and prediction of *noise-free* MR-guided beamlet dose from an extremely noisy (and cheap to simulate) version of the MC beamlet dose for the given geometry. Additionally, we show that our model performs well in previously unseen patient geometries for a given anatomical region such as the head and neck, supporting our expectation of its generalizability for clinical use. We further note that while our model contributes a significant level of acceleration to the task of very-large-scale (VLS) dose calculation, it remains fully compatible with existing MC acceleration techniques such as GPU-based simulation and variance reduction, reinforcing its promise for clinical application.

## 2    Methods

We present a novel technique for accelerated calculation of X-ray beamlet dose from highly under sampled (noisy) Monte Carlo simulation. Our model incorporates the widely successful U-NET CNN architecture to learn the actual dose distribution of an X-ray beamlet, including perturbations resulting from EREs in the presence of an MR-induced magnetic field.

Our model is composed of three independent UNET branches, each with 4 hierarchical levels, that learn a latent representation of each of 3 input channels: under sampled dose, MC X-ray fluence, and CT geometry. Channel-specific latent representations are mixed in a series of fully convolutional layers which preserve the original data dimensionality and produce a prediction of the residual between the input (noisy) and ground-truth dose. Adding the residual and input dose gives the predicted noise-free dose. A summary of the network architecture is shown in figure 1.

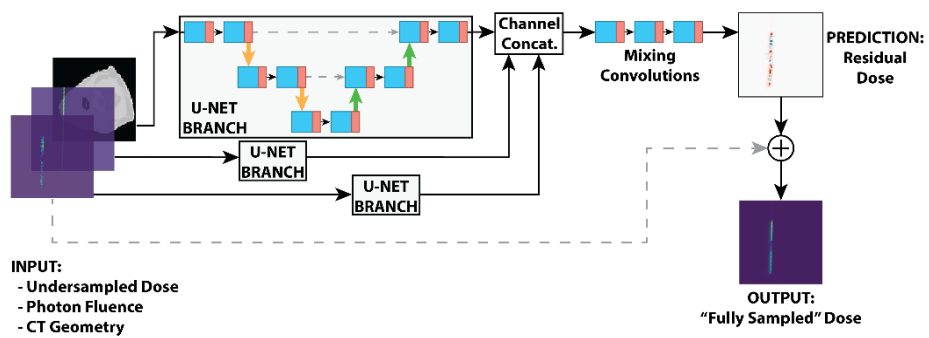

**Fig. 1.** Monte Carlo dose prediction network architecture. Parallel UNET branches process each input channel independently. Concatenation and mixing of latent representations produces predicted residual dose. Residual and under sampled doses are summed, giving prediction of fully sampled dose.

### 2.1 Monte Carlo Dose Simulation

A general purpose, CPU-based Monte Carlo particle simulation toolkit, Geant4 v10.4 [11-12], was used obtain the under sampled and fully sampled beamlet dose distributions as well as the X-ray fluence for each beamlet configuration. A single instance of the fully sampled dose was simulated by tracking 18 million X-rays from a point source 100cm away from the beamlet's isocenter in a uniformly diverging square field. Ten under sampled doses were additionally simulated by instead tracking 500 X-rays each in the same manner. Each beamlet was modeled with an identical histogram-based energy distribution matching that of a clinical 6MV Bremsstrahlung spectrum. To understand the applicability of our approach to MRgRT, we configured a static 1.5T magnetic field, oriented in parallel to the rotation axis of the X-ray source around the treatment isocenter; this geometry matches that of existing MRgRT treatment devices such as the Elekta Unity©.

To standardize the amount of noise present in the under sampled MC dose distributions, we incrementally simulated beamlet dose for 50 randomly selected beamlets in the testing dataset, monitoring the normalized mean absolute error (NMAE) compared with the fully sampled dose until it reached a threshold of 25%. For the fully sampled dose, we selected an average statistical uncertainty during MC simulation of less than 0.1% as the threshold. To maintain these average qualities of dose, the under sampled inputs and fully sampled ground truths were simulated using 500 and 18 million X-rays as described earlier in this section.

### 2.2 Dataset Construction

Beamlets configurations consisted of beam azimuth (gantry angle), isocenter coordinates, and beamlet position within the beam. The parameters of the beamlet configuration were selected randomly to ensure diversity in both the training and testing datasets. Ten head and neck (H&N) CT volumes were retrospectively collected from UCLA's database of radiation treatments and resampled to have an isotropic voxel size of 2.5mm$^3$. Six of the patients were reserved for training and the remaining four for unbiased testing of the trained model. We are careful to test on patients that are previously unseen during the training process so an unbiased evaluation of the model generalizability to new patients can be reported. For each *training* and *testing* patient, an average of 865 and 415 beamlet configurations were randomly sampled, respectively.

A single data example was created by pairing each three-channel input with the fully sampled dose for a specific beamlet configuration. To augment the dataset with extra data examples, rather than randomly generate additional beamlet configurations and perform additional, and expensive, MC simulation of the fully sampled (ground-truth) dose, we recognized that each under sampled simulation of dose is an independent and identically distributed (IID) stochastic observation of the fully sampled dose. This allowed us to pair a single fully sampled dose with multiple (currently 10) independent under sampled inputs. This augmentation technique is like the addition of zero-mean gaussian noise used more commonly in natural image domains, except that we can sample directly from the true noise model by use of MC simulation. After augmentation, our training and testing datasets contained 155,940 and 49,770 examples, respectively.

Our model was trained for 150 epochs (~183,000 iterations) in a data-parallel manner across four NVIDIA GTX Titan X graphics processing units (GPUs). Training time was approximately 18 hours, though the greatest reduction to the loss function was seen after just a few hours. Batch normalization and ReLU operations were used between each convolutional layer.

### 2.3 Experiment Design

To assess the accuracy of the predicted beamlet dose results, we computed the NMAE across every voxel of every beamlet in the testing dataset. To provide physical meaning to this metric, each voxel of the predicted beamlet dose was normalized to the corresponding beamlet-maximum dose, obtained from the fully sampled MC dose volume. We also computed spatial gamma index maps, which indicate the dosimetric accuracy of voxels by combining the dose difference and distance-to-agreement metrics, for each of a pre-determined set of gamma criteria. Readers are referred to [13] for a complete description of the gamma index. Voxels with a gamma index of less than or equal to 1.0, are regarded as *passing* the gamma test, while those with indices in excess of 1.0 are failing, which generally indicate regions of degraded dosimetric accuracy. Our results show *passing* voxels in blue and failing voxels in red, with white indicating the division between the two classifications. Gamma maps are provided for the 0.2%/0.2mm, 0.5%/0.5mm, and 1%/1mm gamma criteria. In our reporting of the results, the NMAE was masked to reduce the bias of less-important voxels with very-low dose. Our masking operation excludes those voxels having both ground truth and predicted normalized dose under 10%, which ensures that both the actual dose and any possible false predictions of dose are low enough to be ignored in most cases.

## 3 Results

In our experiment, where the accuracy of the under sampled MC dose and the deep model predicted dose were compared with respect to the ground truth, a NMAE of 25.7% before prediction and 0.106% after were observed. Prediction time for a single beamlet was approximately 220ms, including both the MC simulation and network prediction steps, while the time to produce a single fully sampled beamlet dose was approximately 380 seconds on average. Figure 2 shows the under sampled (input), network predicted, and fully sampled (ground truth) dose for a single beamlet passing through a large air cavity within the patient's mouth, where EREs are expected and observed. Figure 3 additionally shows the gamma index maps for the beamlet shown in figure 2. Darks blue voxels indicate those that easily pass the gamma test for the imposed criteria (index much less than 1.0). Dark red voxels conversely indicate dramatic failure by the gamma criteria (index much greater than 1.0. Lighter shades of each color, and white indicate voxels that lie near the threshold with index value equal to 1.0.

.........6

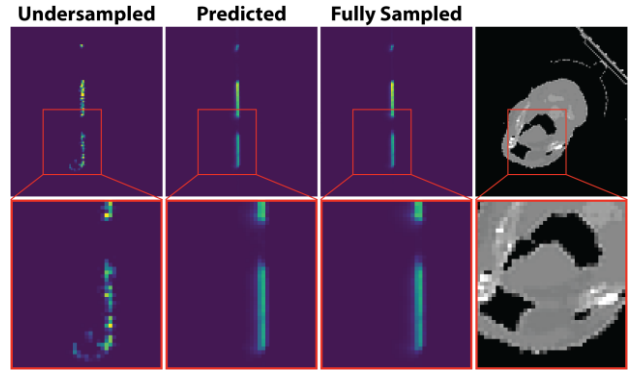

**Fig. 2.** Comparison of under sampled, predicted, and fully sampled (ground truth) dose for one beamlet. Bottom row shows close-up of soft tissue-air interfaces where EREs are visible.

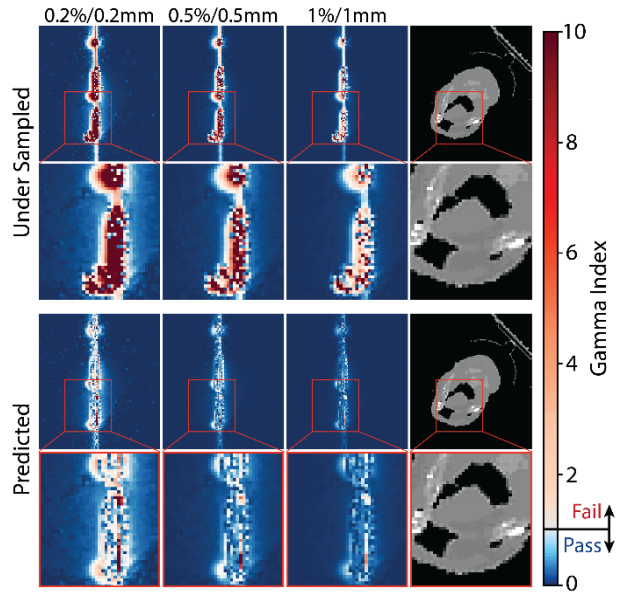

**Fig. 3.** Gamma maps for one under sampled and predicted beamlet dose distribution compared to the ground truth (fully sampled dose). Red voxels indicate large disagreement while white and blue indicate passing for the referenced gamma criterion. Close-up views given under each.

## 4   Discussion

We observe from the analysis of dosimetric accuracy between the under sampled dose and the deep network prediction that a substantial improvement in the beamlet dose accuracy is achieved despite imparting less than 200ms for the additional prediction step. Indicated by the reduction of NMAE for the testing dataset, the accuracy improvement between the under sampled dose and the predicted dose is greater than two orders of magnitude. Without a dedicated analysis of the resulting effects on the treatment planning process, it is difficult to conclude from this study whether the observed accuracy is sufficient for clinical use. However, from the conclusions drawn in [9] we show that our dose prediction model outperforms existing denoising methods with NMAE below 0.2% (improvement ratio of 242) compared to the best performing method of [4] achieving an improvement ratio of only 4.5, corresponding to a NMAE of approximately 1.21% (MSE improvement ratio of 4.5 for the 6.6% uncertainty input) for the H&N evaluation. This improvement is evident despite starting with much noisier dose inputs (input NMAE of more than 25% in our case, compared with up to 6.6% MC uncertainty selected to evaluate the methods of [4] in [9]).

Furthermore, investigating the predicted dose distribution in figure 2 and the corresponding gamma index maps in figure 3 clearly show the advantage of our deep learning-based approach in both the global denoising and the local ERE prediction tasks. For example, the under sampled dose in figure 2 displays a *dose loop* which is commonly observed in noisy MR-guide MC results but is not representative of the expectation obtained by full sampling to a low uncertainty. For these situations, where local filtering approaches tend to fail to distinguish this low probability stochastic event from the true beamlet structure, our model can disambiguate the two and harness the information to produce a more realistic prediction. Moreover, the qualitative differences in the gamma maps of figure 3 clearly demonstrate the global predictive performance of our model, where the fraction of red voxels is substantially reduced between the under sampled and predicted dose distributions.

Like the denoising methods presented in [4-8], our model also benefits from batched evaluation for both the MC simulation and especially the GPU-based model prediction steps. The runtimes reported in section 3 were limited to computation of a single beamlet dose distribution without including the benefits of batched processing. With even a modest availability of GPU hardware and GPU-enabled MC simulation tools, we expect that parallel processing will greatly improve the average per-beamlet processing time well beyond that which is required of online adaptive MRgRT. Further investigation of the limits of acceleration that can be achieved and the benefits to the actual process of treatment plan optimization using predicted beamlet dose distributions are planned for future work.

## 5      Conclusions

We have demonstrated the success of our novel deep learning-based approach to beamlet-scale Monte Carlo dose denoising in terms of the computational time and accuracy improvements. Our technique differs from existing attempts at MC dose denoising in that it: has been evaluated for use in MRgRT where EREs induce local perturbations to the simpler no-magnetic-field X-ray dose distribution, is applicable to substantially noisier dose input resulting from fewer MC-simulated particles, and benefits from efficient deep CNN prediction while maintaining compatibility with existing MC acceleration techniques. By testing our model performance with patient geometries that were not used during model training, our method shows generalizability to new patients, and normalized mean absolute beamlet dose errors of 0.106% on average, compared with the 25.7% error observed by directly using the under sampled dose. This performance is demonstrated while reducing the dose calculation time by over two orders of magnitude compared with fully sampled MC beamlet dose. Our method shows promise in enabling clinical use of adaptive online MRgRT for automatically planned treatments.